# Theoretical studies on the impact of point defect on the structures of different uranium silicides

Miao He[a,b], Shiyu Du[b*], Heming He[c], Jiajian Lang[b], Zhen Liu[b], Qing Huang[b], Cheng-Te Lin[d], Ruifeng Zhang[e] and Dejun Wang[a*]

[a]School of Electronic Science and Technology, Faculty of Electronic Information and Electrical Engineering, Dalian University of Technology, Dalian, 116024, China

[b]Engineering Laboratory of Specialty Fibers and Nuclear Energy Materials, Ningbo Institute of Materials Technology and Engineering, Chinese Academy of Sciences, Ningbo, Zhejiang, China

[c]State Nuclear Power Research Institute, Beijing, 100029, China

[d]Key Laboratory of Marine Materials and Related Technologies, Zhejiang Key Laboratory of Marine Materials and Protective Technologies, Ningbo Institute of Materials Technology and Engineering, Chinese Academy of Sciences, Ningbo 315201, China

[e]School of Materials Science and Engineering, Beihang University, Beijing 100191, P. R. China

## ABSTRACT

The structures, point defects and impacts of fission products for $U_3Si$ (β-$U_3Si$ and γ-$U_3Si$) and $USi_2$ (α-$USi_2$ and β-$USi_2$) are studied by first-principles calculations. The lattice parameters of $U_3Si$ and $USi_2$ are calculated and the stability of different types of point defects is predicted by their formation energies. The results show that silicon vacancies are more prone to be produced than uranium vacancies in β-$USi_2$ matrix, while uranium vacancies are the most stable defects of other three types of crystallographic structures. The most favorable sites of fission products (strontium, barium, cerium and neodymium) are determined in this work as well. By calculating incorporation energies of fission products, we demonstrate that the uranium site is the most favored for all the fissions products. Comparing the structural changes influenced by different fission products, it is also found that the highest volume change is caused by barium interstitials. According to the current data, rare earth elements cerium and neodymium are found to be more stable than alkaline earth metals strontium and barium in a given nuclear matrix. Finally, it is also determined that in $USi_2$ crystal lattice fission products tend to be stabilized in uranium substitution sites, while they are likely to form precipitates from the $U_3Si$ matrix. It is expected that this work may provide new insight into the mechanism for structural

*Corresponding authors. E-mail address: dushiyu@nimte.ac.cn; dwang121@dlut.edu.cn

evolutions of silicide nuclear fuels in a reactor.

1. Introduction

The uranium-silicon binary system has been studied for years due to its potential to be used in the next generation nuclear fuels. The system contains several compounds such as $U_3Si$, $U_3Si_2$, $U_5Si_4$, $USi$, $USi_3$, $USi_2$, $USi_{1.88}$ and $U_3Si_5$($USi_{1.67}$). $U_3Si$ is featured by the high actinide density (14.7 $g/cm^3$) [1] and the better thermal conductivity (30 W/(m·K) at 800 ℃) [2] relative to $UO_2$ (3.4 W/(m·K) at 800 ℃ [3]. The studies on the properties and behaviors of $U_3Si$ have been reported in the literatures. For example, White *et al.* have studied the thermophysical properties of $U_3Si$ from the room temperature to 1150 K [4]. Middleburgh *et al.* have investigated the solubility of Xe and Zr into both the crystalline and amorphous of $U_3Si$ [5]. But there are still difficulties for $U_3Si$ to be applied as a nuclear fuel including its swelling. Hastings *et al.* found that volume changes in irradiated $U_3Si$ are strongly temperature-dependent [6]. $U_3Si$ have three different phases，the α-$U_3Si$ (space group Fmmm) phase forms below -153 ℃ [7], the β-$U_3Si$ (space group I4/mmm) phase forms in the temperature range of -153 to 762 ℃ [8], and β-$U_3Si$ will transform into γ-$U_3Si$ (space group Pm3m) which has $Cu_3Au$ type structure at temperatures above 780 ℃ [9][10]. As another uranium silicide, the investigations on the properties of $USi_2$ have also been performed. The tetragonal α-$USi_2$ is of the $ThSi_2$ type (space group I4$_1$/amd) and essentially the same as that of $USi_{1.88}$, except that it has no deficiency of Si and this compound is metastable at moderate temperatures (< 450 ℃) [9][11][13]. Sasa *et al.* [13] synthesized α-$USi_2$ by leaching excess U from $USi_{1.88}$ in 1:1 HCl solution and the α-$USi_2$ undergoes a complete disproportionation to $USi_{1.88}$ and $USi_3$ when heated (350 ℃) in an evacuated sealed glass tube. A uranium-silicon phase with the $AlB_2$-type (space group P6/mmm) structure, described by Zachariasen *et al.* [12] as stoichiometric β-$USi_2$ and suggested by Kaufmann *et al.* [14] to be deficient in silicon was then shown by Brown *et al.* [15] to be $USi_{1.67}$. The experiment of Brown *et al.* [16] showed that hexagonal β-$USi_2$, prepared by reaction of elements in liquid bismuth, is stable only up to 450 ℃. The corresponding diagrams of these crystal structures are plotted in Fig. 1.

Since both $U_3Si$ and $USi_2$ compounds may be utilized in a nuclear reactor, it is crucial to understand their serving behaviors. As the fission reaction proceeds, fission products may precipitate or agglomerate to form bubbles or precipitates that induce fuel swelling, which can significantly influence the performance of the fuels. Sr is considered as a tracer for spent fuel dissolution and Ba largely contributes to reactor residual heat[17], so the studies on these two fission products are important for the evaluation of the reactor safety. Lanthanides are common fission products as well. As two representative lanthanide products, Nd and Ce are also investigated in this work. Namely, we focus on the stability and impact of fission products Sr, Ba, Nd and Ce in the $U_3Si$ and $USi_2$ matrices. This paper is organized as follows. In Section 2, the computational methods used in this work are briefly presented. In Section 3, the structures and bulk modulus calculated by means of density functional theory (DFT) are compared to experimental data and previous theoretical predictions. Point defects and their influence on crystal lattices are discussed in Section 3 as well. In Section 4, the results on the stability of fission products Sr, Ba, Nd and Ce in the $U_3Si$ and $USi_2$ are reported and their stability in different matrices are discussed. Finally, in Section 5, the results from this study are summarized.

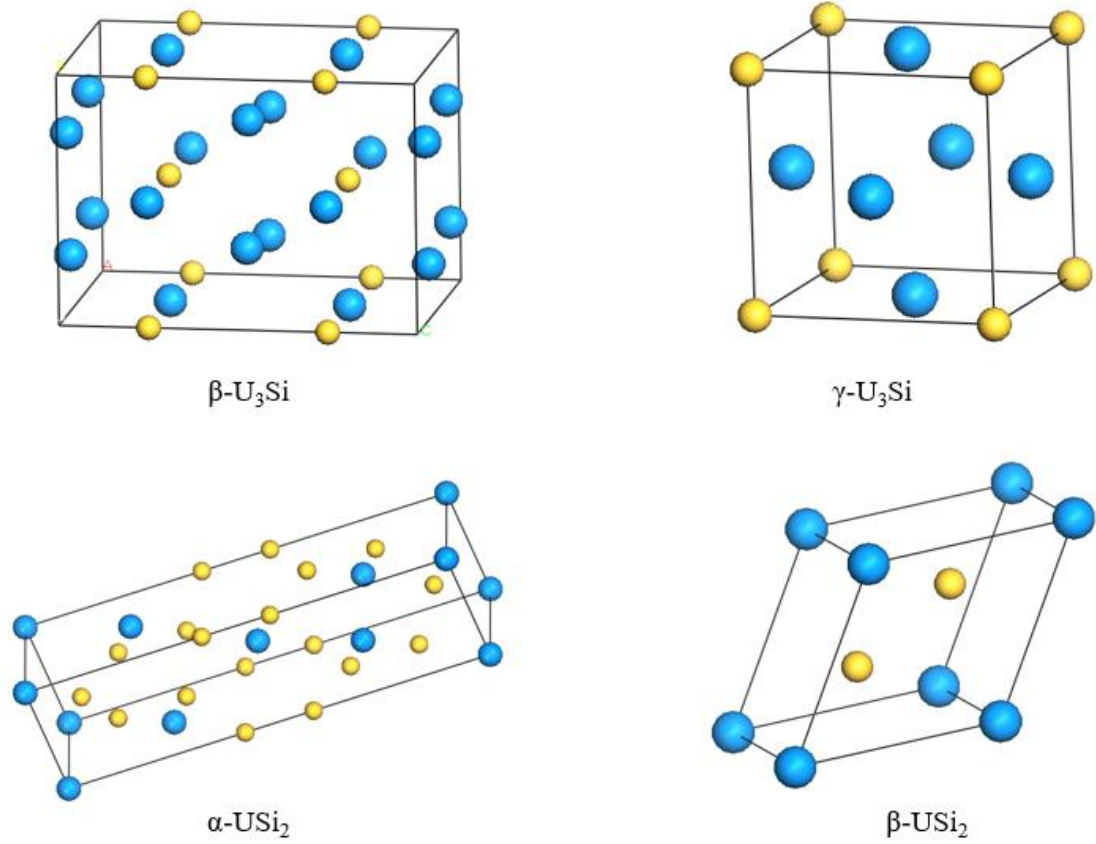

**Fig. 1.** The crystal structure of $U_3Si$ and $USi_2$. The big blue and small yellow filled circles are uranium and silicon atoms, respectively.

## 2. Calculation methods

In this work, the DFT calculations are performed using plane-wave ultrasoft pseudopotential [18][19] as implemented in the Cambridge Serial Total-Energy Package (CASTEP) [20][21]. The exchange and correlation interactions are taken into account with generalized gradient approximation (GGA) as parametrized by Perdew-Burke-Ernzerhof (PBE) [22]. In the calculations, the configurations $6s^2 6p^6 5f^3 6d^1 7s^2$ for uranium and $3s^2 3p^2$ for silicon are adopted to model valence electrons. The SCF convergence threshold is set as $5 \times 10^{-7}$ eV/atom. Geometry optimizations were carried by the Broyden-Fletcher-Goldfarb-Shanno (BFGS) [23] method. The convergence thresholds of geometry optimization are chosen to be $5 \times 10^{-6}$ eV/atom for energy change, and convergence thresholds are 0.01 eV/Å for the maximum force, 0.02 GPa for the maximum stress and $5 \times 10^{-4}$ for the maximum displacement. The calculations are carried out with the approximation of spin non-polarization. In order to investigate lattice perturbations caused by point defects and fission products, both atomic coordinate and volume relaxations of the supercells are taken into account in all calculations. But it is worth noting that the defect formation energies and incorporation energies with volume relaxation have been found to be in reasonable agreement with the constant volume calculations [24] for UC fuel systems. The differences for point defects energies are below 0.1 eV, except that the uranium interstitial may have a deviation of approximately 0.3 eV as shown by Freyss *et al* [24]. For modeling point defects with and without fission products in different phases of uranium silicide, 64-atom supercell (consisting of $2\times2\times1$ primitive unit cells) for β-$U_3Si$, 32-atom supercell (consisting of $2\times2\times2$ primitive unit cells) for γ-$U_3Si$, 48-atom supercell (consisting of $2\times2\times1$ primitive unit cells) for tetragonal $USi_2$ (α-$USi_2$), and 24-atom supercell (consisting of $2\times2\times2$ primitive unit cells) for hexagonal $USi_2$ (β-$USi_2$) are used. A $4\times4\times4$ Monkhorst-Pack [25] k-point mesh and 350 eV cut-off energy for the plane-wave expansion of the electron basis are chosen.

## 3. Bulk properties

Since the difference in exchange-correlation functionals may have impact on the calculation results, the calculations of bulk properties have been performed as the benchmark. **Table 1** lists the lattice constants and bulk modulus calculated with GGA-PBE in our work and GGA-PW91 by Yang *et al.* [26] as well as the experimental data. Here bulk moduli are predicted by the calculated elastic constants [27][28]. The lattice parameters obtained for β-$U_3Si$, γ-$U_3Si$, α-$USi_2$ and β-$USi_2$ are well described by GGA-PBE calculations and the relative error of lattice constants to the experimental values are 0.08%, 1.40%, 1.47% and 0.09%, respectively. But as for the bulk moduli, the calculation results by the GGA-PW91 functional are evidently higher than the corresponding ones with the GGA-PBE method. By comparison with experimental data for $U_3Si$, calculation results of bulk moduli with the GGA-PBE exchange correlation potential appear to agree better with the measurements.

**Table 1** Computed and experimental values of lattice constants (a) and bulk modulus (B).

| Phase | Space group | | | a(Å) | c/a | B(GPa) |
|---|---|---|---|---|---|---|
| β-$U_3Si$ | I4/mcm | This work | GGA-PBE | 6.030 | 1.423 | 119.4 |
| | | Yang | GGA-PW91 | 6.037 | 1.418 | 134.1 |
| | | Exp. | | 6.035 [29] | 1.440 | 101.8 [30] |
| γ-$U_3Si$ | pm3m | This work | GGA-PBE | 4.285 | 1 | 126.7 |
| | | Yang | GGA-PW91 | 4.281 | 1 | 133.1 |
| | | Exp. | | 4.346 [31] | 1 | 118.3 [30] |
| α-$USi_2$ | $I4_1/amd$ | This work | GGA-PBE | 3.869 | 3.714 | 105.6 |
| | | Yang | GGA-PW91 | 3.867 | 3.711 | 116.2 |
| | | Exp. | | 3.922 [13] | 3.610 | |
| β-$USi_2$ | P6/mmm | This work | GGA-PBE | 4.024 | 0.937 | 101.5 |
| | | Yang | GGA-PW91 | 4.038 | 0.936 | 114.8 |
| | | Exp. | | 4.028[16] | 0.956 | |

The formation energies of the compounds calculated by the DFT method can provide insight into the stability of the compounds at low temperatures. The formation energies of the uranium silicides can be calculated by [24]

$$E^F(U_PSi_q) = \frac{E(U_pSi_q) - pE(U) - qE(Si)}{p+q} \quad (1)$$

Here $E^F(U_pSi_q)$ is the formation energy of the compounds $U_pSi_q$. $E(U_pSi_q)$, $E(U)$ and $E(Si)$ are the total energies of $U_pSi_q$, α-U and silicon of diamond structure type calculated using DFT, respectively. **Table 2** gives the calculation results in our work and comparison to the experimental data. O' Hare et al. [32] and Gross et al. [33] have measured the enthalpies of formation for some U-Si compounds using calorimetry. The current results suggest that β-$U_3Si$ are more favorable to be formed than γ-$U_3Si$. It agrees with the facts that the transition temperature between these phases of $U_3Si$ is as high as 762~780 ℃ [7][8][11] and that β-$U_3Si$ can be stable at low temperatures. **Table 2** also shows that both α-$USi_2$ and β-$USi_2$ have high negative formation energies, consistent with the fact that those two compound can be synthesized at low temperature (below 450 ℃) [13][16]. By comparison to experimental enthalpies of formation at 298K, the calculated formation energies exhibit reasonable accuracy, especially for U$_3$Si with an error less than 2.5%, though it is worth accentuating that formation energies are calculated at 0 K here. Herein the results also suggest that GGA-PBE approximation may be satisfactory in reproducing bulk properties as well as the energies of U$_3$Si and USi$_2$.

**Table 2**

Calculated formation energies (eV/atom) of uranium-silicon and experimental enthalpies of formation at 298K.

| Phase | Formation energies (eV/atom) | Enthalpies of formation (eV/atom) |
| --- | --- | --- |
| β-U$_3$Si | -0.278 | -0.271 [32] |
| γ-U$_3$Si | -0.251 | |
| α-USi$_2$ | -0.365 | -0.451 [33] |
| β-USi$_2$ | -0.358 | |

**4. Point defect**

Point defects in nuclear fuels may have significant effect on fuel performances. These defects can provide accommodation sites for fission products and change their diffusion kinetics in nuclear fuel [34]. In this paper, point defects such as uranium/silicon vacancies, uranium/silicon interstitials and uranium/silicon Frenkel pairs are investigated. To study the stability of various defects and assess their influence on crystal structure, we calculate their formation energies and relative volume changes of the lattices caused by point defects. The formation energies of different point defect are calculated by the following expressions:

Formation energy of a vacancy

$$E_{Vx}^{F} = E_{Vx}^{N-1} - E^{N} + E_{x} \qquad (2)$$

Formation energy of an interstitial

$$E_{Ix}^{F} = E_{Ix}^{N+1} - E^{N} - E_{x} \qquad (3)$$

Formation energy of a Frenkel pair

$$E_{FPx}^{F} = E_{FPx}^{N} - E^{N} \qquad (4)$$

Where $E_{Vx}^{F}$, $E_{Ix}^{F}$, and $E_{FPx}^{F}$ are the formation energy of one vacancy x (x = uranium or silicon), one interstitial x and the Frenkel pair of x, respectively. $E_{Vx}^{N-1}$, $E_{Ix}^{N+1}$ and $E_{FPx}^{N}$ are the energies of a uranium silicide supercell containing vacancy x, interstitial x and Frenkel pair of x, respectively. $E^{N}$ is the energy of the uranium silicide supercell without defects; $E_{x}$ is the energy of an "x" in its reference state (α-U crystal or Si of diamond structure type). The energies calculated here are all obtained by the GGA-PBE approximation.

The formation energies for each type of defects in $U_3Si$ and $USi_2$ are given in **Table 4**. Here, the uranium vacancies in those matrices are referred to as Vac U. For β-$U_3$Si, there are two types of U atoms. As shown in Fig. 2a, only uranium atoms are present in Layer 1 and the corresponding vacancy is defined as Vac U, while there are both uranium atoms and silicon atoms in layer 2 and Vac U ´in **Table 4** indicates the uranium vacancy in this layer. Silicon atoms in β-$U_3$Si, γ-$U_3$Si, α-$USi_2$ and β-$USi_2$ are symmetric and their vacancy are all termed as Vac Si. As for interstitial defects, the interstitial structures in different lattices are shown in Fig.2b~Fig.2e with the locations of interstitial defects indicated. Frenkel pairs which are formed by an interstitial atom and a vacancy of the same kind are also investigated. The vacancy and interstitial which compose Frenkel pairs are sited in one supercell and the energies of Frenkel pairs are listed in the column of FPX (X=U, U', Si) in **Table 4**.

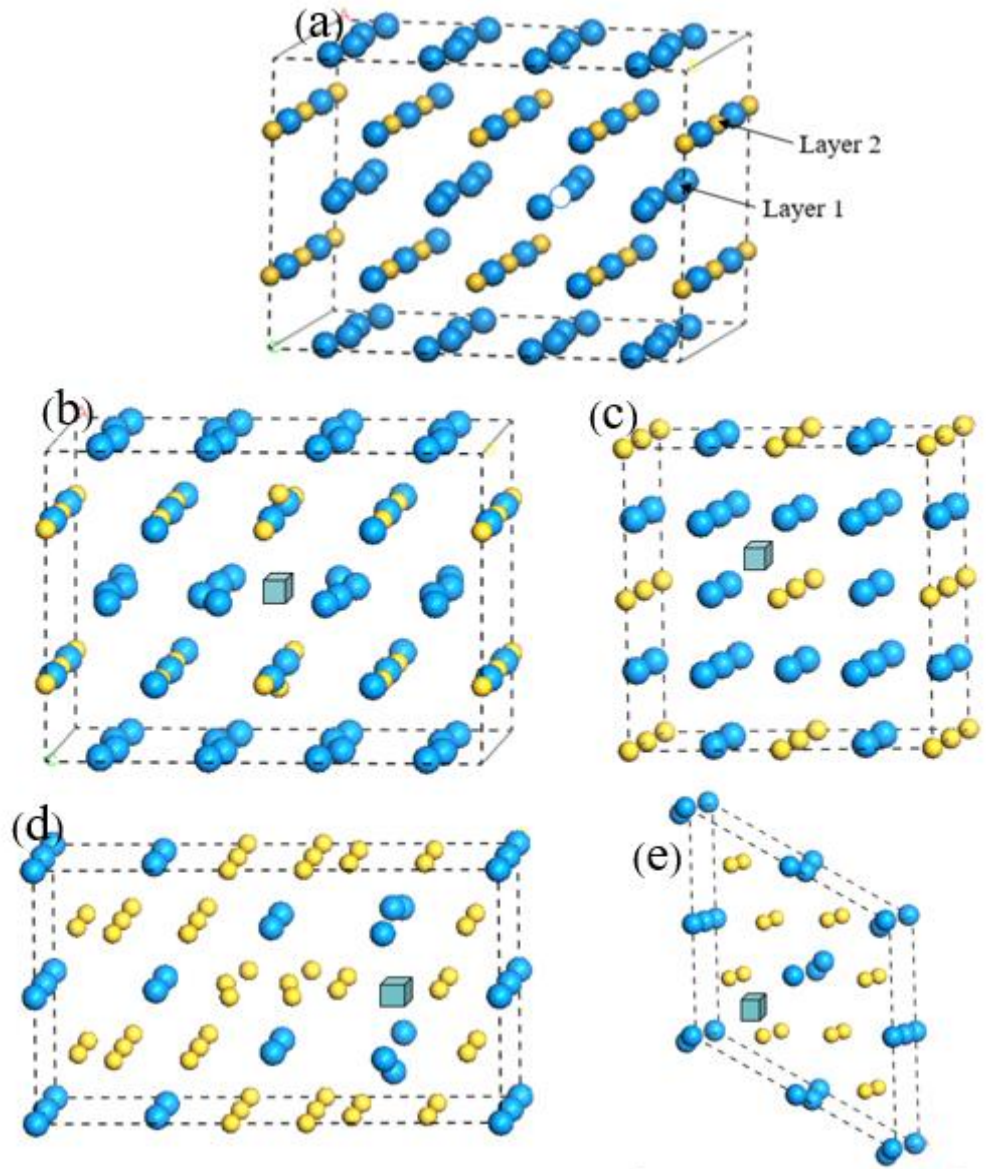

**Fig.2.** Schematic of some different types of defects in uranium-silicon. The big blue and small yellow filled circles are uranium and silicon atoms, respectively. The blue unfilled circles are uranium vacancy. The blue squares are interstitial sites. The five plots are for (a) uranium vacancy in Layer 1 of β-$U_3Si$ (b) interstitial site near the coordinate (0.5, 0.5, 0.5) in β-$U_3Si$ (c) interstitial site near the coordinate (0.5, 0.375, 0.625) in γ-$U_3Si$ (d) interstitial site near the coordinate (0.5, 0.5, 0.25) in α-$USi_2$ (e) interstitial site near the coordinate (0.667, 0.333, 0.5) in β-$USi_2$

**Table 4** Formation energies (eV) of different point defects in different matrices.

| Formation energy (eV) | Vac U | Vac U´ | Vac Si | Int U | Int Si | FP U | FP U´ | FP Si |
|---|---|---|---|---|---|---|---|---|
| β-U$_3$Si | 2.46 | 2.76 | 3.74 | 3.82 | 2.69 | 5.64 | 5.95 | 5.78 |
| γ-U$_3$Si | 2.07 | - | 3.59 | 3.98 | 3.54 | 5.26 | - | 6.84 |
| α-USi$_2$ | 2.76 | - | 3.74 | 2.91 | 4.90 | 4.97 | - | 7.92 |
| β-USi$_2$ | 3.36 | - | 1.52 | 2.67 | 2.92 | 5.15 | - | 4.23 |

From the table, it can be seen that the lowest formation energies belong to the silicon vacancies in β-USi$_2$, suggesting a tendency of formation of silicon vacancies in hypo-stoichiometric USi$_{2-x}$. Thus, the current calculation results can partially explain the reason that non-stoichiometric U$_3$Si$_5$ is substantially the supercell of β-USi$_2$ (U$_3$Si$_6$) with silicon vacancies. In U$_3$Si (both β-U$_3$Si and γ-U$_3$Si) and α-USi$_2$ supercells, uranium vacancies show lower formation energies than other defects. This indicates that U$_3$Si and α-USi$_2$ may generate more uranium vacancies than other defects under the same condition. With respect to the interstitial defects, it is found that silicon interstitials are more stable than uranium interstitials in U$_3$Si which may be caused by the smaller size of Si atom. On the contrary, uranium interstitials have a larger stability than silicon interstitials in USi$_2$ (both α-USi$_2$ and β-USi$_2$), showing the accommodation of uranium atom in the lattice of USi$_2$ is more favored than that of U$_3$Si. It is also interesting to point out uranium vacancies prefer to locate in layer 1 rather than layer 2 in β-U$_3$Si probably because no U-Si bonding interaction is changed when vacancies form in Layer 1. As for the Frenkel defect pairs, since they are constituted of both vacancies and interstitial defects and require more energies to form than other types of defects, one can readily find that all Frenkel defects have large formation energies which are above 5 eV in all matrices except for U of α-USi$_2$ and Si of β-USi$_2$. Moreover, the formation energies of Frenkel pairs are not identical to the sum of those calculated for isolated vacancy and interstitial point defects due to the limited size of supercells, which means the vacancy and interstitial in one supercell are actually bound and have mutual attraction. According to the formation energies of bound Frenkel pairs listed in **Table 4**, the "attraction" between the vacancy and the interstitial defects are small (0.29 and 0.21 eV, respectively) for Si Frenkels of γ-$U_3Si$ and β-$USi_2$ but large (>0.5 eV) for those of β-$U_3Si$ and α-$USi_2$. For bound Frenkel pairs, the attraction between vacancy and interstitial defect is smaller when their distance is longer (theoretically, it goes to zero when the distances are long enough). Interestingly, the size of the supercells of γ-$U_3Si$ and β-$USi_2$ adopted in this work are smaller than β-$U_3Si$ and α-$USi_2$, respectively. This strongly suggests that the defects of silicon atoms in the γ-$U_3Si$ and β-$USi_2$ matrices may have a shorter correlation length than β-$U_3Si$ and α-$USi_2$. Additionally, according to the table, silicon Frenkel defects have relatively larger formation energies than uranium ones except for β-$USi_2$.

It is well known that point defects may also induce changes in the bulk properties such as densities of nuclear fuels. **Table 5** represents relative volume variation ($\Delta V/V$) of the crystalline structures of uranium silicides induced by defects. The $\Delta V/V$ values are determined by comparing the optimized lattice volume of the supercells with one defect with that of the perfect supercell structure. It should be kept in mind that relative volume variation is a function of defect density related to the size of the adopted supercell, however, it can still qualitatively reflect the perturbations of structure induced by different types of defects. One can see that for all systems interstitial defects always cause a swelling of crystal lattice. On the other hand, vacancies can diminish the volume of the crystal though the relative variation is smaller. These feature are similar to that of $UO_2$ and UC [24][36], but are different from that of $U_3Si_2$, in which the X (X=U, Si) vacancy also causes swelling [37]. For any given compound, uranium interstitials appear to induce more volume change than silicon interstitials probably because uranium has a larger atomic radius than that of silicon. Especially, it is worthy of mentioning that there is little difference (not exceeding 0.15%), between the relative variation caused by U vacancy and that caused by U´ vacancy in β-$U_3Si$, as well as U Frenkel and U´ Frenkel pairs, which demonstrates the atomic radius is the major factor to determine the swelling effect in β-$U_3Si$. The isolated Frenkel would cause swelling of crystal as listed in **Table 5**, indicating swelling of the volume caused by X interstitial surpasses the volume diminishing effect by X vacancy. Therefore, the diffusion of U or Si atom in a uranium silicide matrix may always induce the lattice swelling.

Table 5 Volume variation (%) of uranium-silicon induced by different point defects.

| Volume variation (%) | Vac U | Vac U´ | Vac Si | Int U | Int Si | FP U | FP U´ | FP Si |
|---|---|---|---|---|---|---|---|---|
| β-U$_3$Si | -0.69 | -0.57 | -0.43 | 1.87 | 1.74 | 1.38 | 1.52 | 1.32 |
| γ-U$_3$Si | -1.21 | - | -0.75 | 5.85 | 4.68 | 3.65 | - | 3.82 |
| α-USi$_2$ | -0.57 | - | -0.43 | 2.77 | 2.61 | 2.10 | - | 5.62 |
| β-USi$_2$ | -0.15 | - | -0.96 | 5.44 | 5.01 | 4.63 | - | 3.96 |

## 5. Fission products

In this section, the stabilities of fission products in different uranium silicide fuels are studied. Various fission products can be generated during a nuclear reaction among which four representative fission products: Sr, Ba, Nd and Ce are chosen in this work. Since the fission products may also affect the uranium silicide crystal structures, the relative volume variation of the structures caused by fission products is predicted. Here, three different sites are considered: 1. the uranium substitution sites in uranium silicide crystals (as mentioned there are two different uranium sites in β-U$_3$Si); 2. the silicon substitution sites; 3. the interstitial sites as show in Fig.2b~e.

Incorporation energies ($E_x^{Inc}$) provide the information on the stability of fission products (x) in the defective nuclear fuel matrix. In this work, $E_x^{Inc}$ of different kinds of fission products in different matrices are studied, which are predicted as energy needed to locate an isolated atom into a pre-existing point defects or an interstitial site, i.e.:

$$E_x^{Inc} = E_x^{Tol} - E^{Tol} - E_x \qquad (5)$$

Here $E_x^{Tol}$ is the energy of one uranium silicide supercell with a fission product, $E^{Tol}$ is the energy of the uranium silicide supercell with a defect, and $E_x$ is the energy of the dissolved atom (Sr, Ba, Nd or Ce) in its reference state.

**Table 6**

Incorporation energies (eV) of strontium, barium, cerium and neodymium at the uranium and silicon substitution sites and interstitial site (Site U, Site U', Site Si and Site Int) in different matrices.

| Incorporation energies (eV) | | Sr | Ba | Ce | Nd |
|---|---|---|---|---|---|
| β-$U_3Si$ | Site U | 0.83 | 1.82 | -0.15 | -2.74 |
| | Site U´ | 0.85 | 1.78 | -0.22 | -2.93 |
| | Site Si | 1.39 | 1.70 | 0.77 | -1.67 |
| | Site Int | 7.98 | 9.01 | 4.90 | 3.09 |
| γ-$U_3Si$ | Site U | 0.89 | 1.17 | -1.68 | -2.43 |
| | Site Si | 1.30 | 1.56 | -1.52 | -1.86 |
| | Site Int | 6.05 | 6.44 | 3.77 | 2.29 |
| α-$USi_2$ | Site U | -5.12 | -4.19 | -7.02 | -7.25 |
| | Site Si | 5.05 | 5.83 | -0.51 | -3.54 |
| | Site Int | 8.82 | 9.35 | 3.91 | 2.41 |
| β-$USi_2$ | Site U | -2.21 | -1.24 | -3.85 | -4.43 |
| | Site Si | 4.27 | 6.45 | 1.39 | -2.54 |
| | Site Int | 8.95 | 9.86 | 4.12 | 2.32 |

The calculation results are displayed in **Table 6**. One can see that the sequence of stability can be queued as (Site U) > (Site Si) > (Site Int) when a fission products and fuel matrix are given. It means that fission products always prefer to be accommodated by the U sites due to the larger space a uranium vacancy provides. Similar to uranium silicide, some fission products also have a high tendency to occupy the U sites instead of the C sites, N sites and O sites in UC, UN and $UO_2$ matrices, respectively [24][38][39][40][41]. However, it is noteworthy that this feature may not be universal for different fission products. For example, Middleburgh *et al.* calculated the incorporation energy of Xe and Zr in β-U$_3$Si and found that Zr seem to prefer Si sites than U sites with the incorporation energies of -1.79 and -1.53 eV, respectively [5].

In terms of different fission products, for all nuclear fuel matrices, rare earth elements are more stably dissolved than alkaline earth metals. Nd is the most stable and Ba is always the most unstable in all three kinds of solution sites with the incorporation energies following $E_{Nd}^{inc} < E_{Ce}^{inc} < E_{Sr}^{inc} < E_{Ba}^{inc}$. The computational results are similar to Bévillon *et al.*'s work [37]. They have calculated incorporation energies of the fission products in UC matrix and concluded that the incorporation energies of fission products follow $E_{Nd}^{inc} < E_{Ce}^{inc} < E_{Ba}^{inc}$. These results are also consistent with the findings in Brillant *et al.*'s and Gupta *et al*'s reports on UO$_2$ [39][40]. According to their data obtained with the GGA+U method, it is found that the incorporation energies of fission products in the same site also show $E_{Ce}^{inc} < E_{Ba}^{inc}$.

According to the results, the incorporation energies of all fission products in U site are always negative in the USi$_2$ matrix, indicating the stability of these fission products in the U site of USi$_2$. As to the U$_3$Si matrix, only incorporation energies of rare earth elements in U site are negative, which probably results from the resemblance of the valence electron configuration between Ce/Nd and uranium. However, they are still less stable than their counterpart in USi$_2$ matrix. As an example, the difference of incorporation energies is as high as 6.87 eV between β-U$_3$Si and α-USi$_2$ for Ce at U site; the difference also exceeds 2 eV between γ-U$_3$Si and β-USi$_2$ for Nd. Hence, fission products prefer to stay in the U sites and are more stable at U site in the USi$_2$ matrix, while, those in high uranium density U$_3$Si matrix have a higher propensity to form precipitates which will induce fuel particle swelling, as confirmed by experiments [42][43][44][45]. For example, Finlay *et al.* found that some fuel candidates with high uranium density such as U$_3$Si exhibit high swelling rates even at low and medium fission densities while the lower density compounds like USi show swelling rates which were significantly reduced and are regarded as stable and acceptable[44].

Hence, our theoretical studies may provide an explanation to the swelling of U$_3$Si pellet, though the complete mechanism may be complicated. But due to concern of the high actinide density and good thermal conductivity, the U$_3$Si compounds may still be chosen as nuclear fuel as USi$_2$. Additionally, it should be noted that the incorporation energies alone only reflect the stability of fission products in the trap site that has been formed, e.g. a preexisting U vacancy. The solution energy (incorporation energy + the trap site formation energy) can be used as a measurement for the solubility. For example, the incorporation energy of Ba in a preexisting U vacancy in β-USi$_2$ matrix is determined to be negative (-1.24 eV), whereas the solution energy of Ba in the U site is positive (2.12 eV). It probably means that Ba atoms are mainly captured by preexisting U vacancies in a β-USi$_2$ crystal.

**Table 7** gives relative volume variation of crystal caused by the fission products. It can be seen that volume of the uranium silicide lattice will increase when there exists a fission product of any type in any site, which is partly due to the lager radii of the Nd, Ce, Sr and Ba atoms than those of the U and Si atoms. For a given site and a given nuclear fuel matrix, e.g. for U site in β-U$_3$Si, the sequence of the relative volume variation caused by different fission products are as follows: ΔV/V(Ba) >ΔV/V(Sr)>ΔV/V(Ce)>ΔV/V(Nd), indicating that alkaline earth metals (Sr and Ba) causes a larger change of crystal volume than rare earth elements (Ce and Nd) in U$_3$Si and USi$_2$, which is consistent with our prediction in incorporation energies. For the elements from the IIA group of the periodic table, one has: ΔV/V(Ba) > ΔV/V(Sr) which can be attributed to the size of the fission products atom (R$_{Ba}$ > R$_{Sr}$). The data ΔV/V(Ce) > ΔV/V(Nd) may result from the fewer *f* electrons of the Ce atom. For a given fission product and a given nuclear fuel matrix, volume variation become larger as the size of the defect decreases and the variation may reach maximum for fission products located in interstitial sites, i.e. ΔV/V(Site U)<ΔV/V(Site Si)<ΔV/V(Site Int). The sequence of swelling also appears to be opposite to that of the stability according to the data in **Table 6**, which means the instability in energetics caused by the fission products is reflected by the apparent lattice swelling. In other words, the volume variation provides a different way for the understanding of stability of the uranium silicide systems in a reactor.

**Table 7**

Volume variation (%) of uranium-silicon induced by strontium, barium, cerium and neodymium at different sites-uranium, silicon substitution sites and interstitial site (Site U, Site Si and Site Int) in different matrices

| Volume variation (%) | | Sr | Ba | Ce | Nd |
|---|---|---|---|---|---|
| β-$U_3Si$ | Site U | 0.84 | 1.13 | 0.73 | 0.31 |
| | Site U´ | 0.71 | 1.09 | 0.69 | 0.33 |
| | Site Si | 1.08 | 1.13 | 1.10 | 0.46 |
| | Site Int | 2.92 | 3.95 | 2.31 | 1.32 |
| γ-$U_3Si$ | Site U | 1.99 | 2.68 | 0.84 | 0.40 |
| | Site Si | 2.37 | 2.40 | 1.42 | 0.99 |
| | Site Int | 7.90 | 8.23 | 5.67 | 4.74 |
| α-$USi_2$ | Site U | 0.56 | 0.83 | 0.35 | 0.27 |
| | Site Si | 2.63 | 3.78 | 2.12 | 1.56 |
| | Site Int | 4.51 | 5.35 | 2.46 | 2.43 |
| β-$USi_2$ | Site U | 2.70 | 4.62 | 1.54 | 1.33 |
| | Site Si | 4.79 | 7.25 | 2.33 | 1.46 |
| | Site Int | 7.83 | 8.56 | 7.47 | 5.34 |

## 6. Conclusion

In this work, the bulk properties, point defects and incorporation energies of fission products in different uranium silicide fuel matrices are investigated using the generalized gradient approximation in the framework of density functional theory. The calculated lattice constants and bulk modulus agree well with the experimental data, which shows that the DFT-GGA method is satisfactory to model the behavior of $U_3Si$ and $USi_2$ in the atomic scale. The study of point defects shows that Si vacancies are easier to be produced than U vacancies in the $\beta$-$USi_2$ matrix, while in other matrices studied, U vacancies are more readily produced than Si vacancies. Interstitials always lead to volume expansion and vacancies diminish the volume of the crystalline structure of $U_3Si$ and $USi_2$. In all matrices, rare earth elements Ce and Nd are more stable than alkaline earth metals Sr and Ba. And alkaline earth metals cause a larger change of crystal volume than rare earth elements. The stability and volume variation of matrix crystal are determined to be dependent on the atomic size of the fission products. The U sites may provide the most stable accommodation for all fission products in $U_3Si$ and $USi_2$. Fission products in $U_3Si$ matrix are more likely to form precipitates than $USi_2$.

**Acknowledgments**

The authors acknowledge the support of the National Key Research and Development Program of China (No. 2016YFB0700100), the Division of Functional Materials and Nanodevices, Ningbo Institute of Materials Technology and Engineering, Chinese Academy of Sciences, the National Natural Science of Foundations of China (Grant Nos. 51372046, 51479037, 91226202 and 91426304), the Major Project of the Ministry of Science and Technology of China (Grant No.2015 ZX06004-001), the Ningbo Municipal Natural Science Foundation (No. 2014A610006), One Thousand Youth Talents plan, ITaP at Purdue University for computing resources and the key technology of nuclear energy, 2014, CAS Interdisciplinary Innovation Team.